\documentclass[journal]{IEEEtran}
\usepackage[utf8]{inputenc}
\usepackage[english]{babel}
\usepackage{siunitx}
\usepackage{textcomp}
\usepackage{graphicx}
\usepackage{amsmath}
\usepackage[dvipsnames]{xcolor}
\DeclareUnicodeCharacter{2212}{+}
\usepackage{booktabs}
\ifCLASSINFOpdf
\else
\fi
\begin{document}
%
\title{Noise adaptive beamforming for linear array photoacoustic imaging}
%
%
%
\author{Souradip Paul,~\IEEEmembership{Graduate Student Member,~IEEE,}
        Subhamoy Mandal,~\IEEEmembership{Senior Member,~IEEE,}\\and~Mayanglambam Suheshkumar Singh$^*$,~\IEEEmembership{Member,~IEEE}
\thanks{S Paul and MS Singh$^*$ are with the School of Physics, Indian Institute of Science Education and Research Thiruvananthapuram, Thiruvananthapuram - 695551, Kerala, India. Contact: (email: suhesh.kumar@iisertvm.ac.in; see:http://www.iisertvm.ac.in/faculty/suhesh.kumar) }
\thanks{S Mandal and is with the Division of Medical Physics in Radiology, German Cancer Research Center (DKFZ), Heidelberg, 69120 Germany, e-mail: s.mandal@ieee.org;}
\thanks{S Mandal and S Paul contributed equally; MS Singh supervised the work.}
\thanks{Manuscript received October 20, 2020; revised XX}}

\maketitle
\begin{abstract}
Delay-and-sum ($DAS$) algorithms are widely used for beamforming in linear array photoacoustic imaging systems and are characterized by fast execution. However, these algorithms suffer from various drawbacks like low resolution, low contrast, high sidelobe artifacts and lack of visual coherence. More recently, adaptive weighting was introduced to improve the reconstruction image quality. Unfortunately, the existing state-of-the-art adaptive beamforming algorithms computationally expensive and do not consider the specific noise characteristics of the acquired ultrasonic signal. In this article, we present a new adaptive weighting factor named the variational coherence factor ($VCF$), which takes into account the noise level variations of radio-frequency data. The proposed technique provides superior results in terms of image resolution, sidelobe reduction, signal-to-noise and contrast level improvement. The quantitative results of the numerical simulations and phantom imaging show that the proposed $VCF$ assisted $DAS$ method leads to  55\% and 25\% improvement in FWHM, 57\% and 32\% improvement in $SNR$, respectively, compared to the state-of-the-art $DAS$-based methods. The results demonstrate that the proposed method can effectively improve the reconstructed image quality and deliver satisfactory imaging performance even with a limited number of sensor elements. The proposed method can potentially reduce the instrumentation cost of the photoacoustic imaging system and contribute toward the clinical translation of the modality.
\end{abstract}

\begin{IEEEkeywords}
Photoacoustic imaging, ultrasonic beamforming, adaptive signal processing
\end{IEEEkeywords}

%
\IEEEpeerreviewmaketitle

\section{Introduction}
\label{sec:introduction}
\IEEEPARstart{P}{hotoacoustic}/optoacoustic imaging is an emerging non-invasive biomedical imaging modality based on photoacoustic (PA) effect \cite{wang2012photoacoustic,manohar2016photoacoustics,singh2020fundamentals,suhesh2020IOPbook}. In PA imaging, electromagnetic (EM) intermittent radiation gets absorbed by the tissue material (chromophores) resulting in local temperature rise causing thermoelastic expansion in the tissue \cite{singh2020fundamentals,suhesh2020IOPbook,beard2011biomedical,roy2019fluid}. The rapid expansion (and subsequent contraction) of tissues generate broad-band ultrasound (US) waves (known as PA waves) that propagate to the tissue boundary. These US waves are recorded using ultrasonic detector to reconstruct PA images \cite{wang2012biomedical}. PA imaging combines the optical spectral contrast with benefits from US detection $-$ more specifically, low acoustic scattering of tissue and thus, high resolution $-$ that enables to characterize structural, functional, physiological and molecular information of deeply seated (living) tissues at an unprecedented resolution \cite{7106682, singh2014elastic, kim2020towards,cai2016detection}. The unique ability of PA imaging to probe biological tissues and their patho-physiological information at an unprecedented imaging depth (with unprecedented spatial resolution) has enabled researchers to expand the horizon of biomedical research and its clinical applications \cite{ermolayev2016simultaneous, piras2009photoacoustic,pai2017accuracy}.  

PA image reconstruction is crucial to recover the distribution of initial pressure waves ($P_0$) from which one can derive the patho-physiological information and its distribution over a pre-specified region of interest \cite{ schoonover2011image}. There are several approaches to image reconstruction that are inspired by imaging geometry, for example, beamforming methods for linear array transducers \cite{ma2019multiple, matrone2014delay, mozaffarzadeh2018enhanced} and tomographic methods for array transducers of circular and hemispherical geometrical shape. In the current work we are primarily focused in improving the beamformed image quality by using a noise adaptive model \cite{wang2014snr, hasegawa2020converting}. Real-time PA image reconstruction is highly impactful in clinical setting \cite{xia2014photoacoustic, jeon2019real}. Delay-and-sum ($DAS$) has emerged as a powerful tool to generate real-time reconstruction images which is equipped by its low complexity and/or simple implementation. It remains as the most commonly used real-time image reconstruction technique in US imaging (but as an upcoming and promising technique in PAI). However, the native $DAS$ reconstruction algorithm suffers from serious drawbacks, namely, poor spatial resolution, low contrast, high level side lobes and intense artifacts \cite{hoelen2000image}. The algorithm is primarily based on calculations of geometrical distance between the transducer element and the target pixel point. The algorithm assumes that the pixels, which are at same distance, have similar PA signal intensity. However, for linear array transducers, the cone-shaped acoustic beam from a PA (point) source directing towards a linear array detector does not scale-up linearly due to heterogeneity in spatial distribution of tissue physical (acoustic) properties, transducer directivity, signal transmissibility and similar confounding factors. These inherent factors create high level side lobes and artifacts in the reconstructed images \cite{ma2019multiple}. 

Several studies have been reported in the past to address the fundamental limitations of $DAS$ beamforming algorithm. $DAS$ beamforming combined with coherence factor ($CF$), modified coherence factor ($MCF$) has been used in PA imaging in order to reduce unwanted off-axial signal contribution to beamforming \cite{mozaffarzadeh2018enhanced}. In 2015, Matrone \emph{et al.} \cite{matrone2014delay} proposed a new beamforming algorithm, named as delay-multiply-and-sum ($DMAS$) beamformer, to overcome the limitations of DAS in US imaging. The improved DMAS algorithm shows better performance in contrast enhancement, reducing side lobes and high resolution but it is challenging to adapt in clinical setting due to its high computational complexity. Mozaffarzadeh \emph{et al.} \cite{mozaffarzadeh2017double} introduced modified and improved version of $DMAS$, called as double stage $DMAS$. However, the technique is computationally more expensive and thus, unable to generate real-time images. Later, real-time $DMAS$ that reduced computational burden was established \cite{jeon2019real}. On the other hand, minimum variance ($MV$) and Eigenspace based minimum variance ($EIBMV$) beamforming combined with $DAS$ and $DMAS$ have also been exploited to further improve the performance of PA image reconstruction \cite{mozaffarzadeh2018linear, mozaffarzadeh2018eigenspace}. Various modifications of MV beamforming have been investigated for reduction in computational complexity. Another beamforming technique, namely, delay-and-standard-deviation ($DASD$) \cite{matrone2014delay}, was applied to enhance the image contrast for some particular cases of US imaging such as needle biopsy or cardiac catheterization. Dopper based Motion Compensation Strategies (MoCo) is proposed to achieve good quality volumetric B-mode images that are minimally affected by motion artifacts \cite{chen2018doppler}. Weighted $DASD$ combined with $DAS$ and $DMAS$ significantly reduces the noise level as well as suppresses the side lobes \cite{Roya_a}. Short-lag spatial coherence beamformer was introduced by Lediju \emph{et al.} \cite{lediju2011short} for contract enhancement in PA imaging. A sparsity based minimum variance was proposed to enhance the contrast of PA images, retaining the significant full-width-half-maximum of minimum variance \cite{shang2018sparsity}. However, the reported techniques improve the quality of the reconstructed PA images compared to non-adaptive ones at the price of higher computational complexity.

In this article, we report a novel weighting function named variational coherence factor ($VCF$), which is fundamentally based on the measurement of mean and standard deviation of the detected delayed signals. It is proposed to enhance the image quality compared to existing beamforming techniques ($DAS$, $DAS+CF$). Furthermore, we implement a simplified accelerated form of this method, which can be expressed as a functional form of conventional $CF$ and theoretical noise model to validate the performance of this weighting method under different noise conditions. The proposed algorithm achieves outstanding performance in PA image reconstruction $-$ more specifically, higher spatial resolution, higher off-axis interference rejection, higher signal-to-noise ratio ($SNR$) and higer contrast ratio ($CR$) $-$ in comparison with the existing $DAS$ and $CF$ based beamforming algorithms. In addition, we present characterization study on robustness in performance of the proposed algorithm under various noise levels (increasing channel $SNR$) and achievable image quality as a function of detector elements ($64$, $128$, and $256$) in comparison with the conventional beamforming methods ($DAS$ and $DAS$ $+$ $CF$). The image quality metrics (like $SNR$, $CR$, $gCNR$) in $VCF$ based method show a better improvement than the conventional method ($DAS$, $DAS+CF$) under different channel noise levels, transducer elements, and transducer frequency responses. The numerical simulation and experimental results demonstrate that our proposed algorithm adaptively increases the robustness in noise reduction compared to the state-of-the-art methods. Therefore, It is a promising technique not only for real-time image reconstruction but also with a limited number of sensor elements. The reasonably good or satisfactory (reconstructed) image quality achievable with a limited number ($ 64$) of sensor elements $-$ against commonly used $128$ $-$ will drastically reduce the cost of PAI system or paves a way for development of affordable PAI system that will surely impart to easy access of this promising imaging system to economically weak countries where clinical infrastructure is limited. Shortly, our present study can impart significantly to change in clinical setting. Rest of the paper is organised as: necessary materials and methods are explained in Sec. \ref{sec:material_methods}, numerical simulation and experimental results are provided in Sec. \ref{sec:simulation} and Sec. \ref{sec:expt_results} respectively, discussions are presented in Sec. \ref{sec:discussion} and conclusion is given in Sec. \ref{sec:conclusion}.    

\section{Materials and Methods}
\label{sec:material_methods}
\subsection{Delay-and-sum ($DAS$)}
In $DAS$ beamforming technique, the recorded time-resolved PA signals are properly aligned by taking into consideration the time delays corresponding to radial distance between the positions of individual sensor elements in a (linear) transducer unit and imaging targets of interest for reconstruction. Then, in order to obtain the reconstructed initial pressure distribution that is displayed as an image, the delayed signals of the individual transducer elements corresponding to pixels in the (discretized) reconstruction image plane are algebraically summed-up. Mathematically, $DAS$ beamforming algorithm can be expressed as: 
\begin{equation}
    S_{DAS} [m,n] = \sum_{i=1}^{N}s_i \big [k + \Delta_i [m,n] \big ],
\end{equation}
where $S_{DAS} [m,n] $ is the output beamformed or reconstructed initial PA signal strength at pixel position $[m,n]$ in the discretized reconstruction $mn$-space; $N$ is the number of sensor elements in the transducer unit; $\Delta_i [m,n]$ is the time delay in time-resolved PA signals being recorded by $i^{\text{th}}$ sensor element (relative to the sensor facing opposite to the source being situated at $[m,n]$); $k$ is the time index of discrete PA signal being recorded by the transducer element facing opposite to the source at $[m,n]$. One can estimate $\Delta_i [m,n]$ from the measurement of acoustic path difference ($\Delta r_i [m,n] $) of PA signal arriving at $i^{\text{th}}$ sensor element with respect to that facing opposite to $[m,n]$, i.e., $\Delta_i [m,n] = \frac{\Delta r_i [m,n]}{c}$ (where $c$ is the speed of sound in the (tissue) medium). From practical aspects, the time-resolved PA signals being received by the individual transducer elements are continuous in nature. This (continuous) time-resolved PA signals are acquired and stored as digital (discretized and quantized) signals in the process of data acquisition. Here, $[.]$ represents discrete time in contrast to $(.)$ that conventionally represents continuous time domain.

\subsection{Coherence factor ($CF$)} 
Coherence factor ($CF$) $-$ which is one of the well celebrated non-linear weighting factors being introduced to suppress side lobe artifacts $-$ is defined as \cite{jeon2019real}: 
\begin{eqnarray}
{\scriptstyle CF[m,n] = \frac{ (\sum_{i=1}^{N}s_i [k + \Delta_i [m,n] ] )^2}{N \sum_{i=1}^{N} s_i^2[k + \Delta_i [m,n] ]} = \frac{<s [k + \Delta [m,n]]>^2}{<s^2[k + \Delta [m,n]]>}.} \label{eq:CF}
\end{eqnarray}
where $<.>$ represents mean or average. The numerator in \ref{eq:CF} represents the energy of the coherent sum obtained in a conventional $DAS$ method, and the denominator is interpreted as the total energy (contain both coherent and incoherent components of the PA signals) \cite{hasegawa2020converting}. The maximum range of the $CF$ is bounded to 1.  

Combining $CF$ with $DAS$, we can express the $CF$ based beamformer as \cite{mozaffarzadeh2018enhanced}:
\begin{equation}
    S_{DAS + CF}[m,n] = S_{DAS}[m,n]\times CF[m,n].
\end{equation}

\subsection{Variational coherence factor ($VCF$)}
Variational coherence factor ($VCF$), that we are introducing in this study, is a functional form of coherence factor and it is fundamentally based on measurement of mean and standard deviation ($\sigma$) of the delayed PA signals detected by individual sensor element. The time delayed PA signals corresponding to a particular position in the reconstruction plane is summed up $-$ which is typically performed by $DAS$ $-$ and the sum is divided by the standard deviation ($\sigma$) of the delayed signals which is a measure of fluctuation or dispersion from the mean. Mathematically, the proposed ($VCF$) weighting factor is defined as:  
\begin{eqnarray}
{\scriptstyle VCF [m,n]}  &=& {\scriptstyle \frac{\frac{1}{N}\sum_{i=1}^{N}s_i [k + \Delta_i [m,n] ] }{\sqrt{\frac{1}{N}\sum_{i=1}^{N}s_i ^2[k + \Delta_i [m,n] ] - (\frac{1}{N}\sum_{i=1}^{N}s_i [k + \Delta_i [m,n] ] )^2}}} \nonumber \\
         &=& {\scriptstyle \frac{<s [k + \Delta [m,n]]>}{\sqrt{<s^2 [k + \Delta [m,n]]> - <s [k + \Delta [m,n]]>^{2}}}}, \nonumber \\
         &=& {\scriptstyle \sqrt{\frac{CF[m, n]}{1-CF[m, n]}}}, \label{eq:VCF_CF}
\end{eqnarray}
which shows that $VCF$ is a function of $CF$. A detailed derivation is given in supplementary document (S). $CF$ evaluates the ratio of coherence component in the received PA signals to the total energy at a given point in the (image) reconstruction space. Therefore, whenever noise level is high in detected PA signal, $CF$ provides a sub optimal performance because noise suppresses the contribution of coherency and thus affects the efficiency of the sidelobe rejection by  $CF$ method \cite{wang2014snr}. In contrast, our proposed method can easily separate out the incoherence contribution from beamformed signals. Eq.\ref{eq:VCF_CF}, It can be shown that denominator of $VCF$ contains only the incoherent part. As the maximum value of $CF$ is limited to $1$, so ($1-CF$) facilitate the contribution of noise or incoherency in the beamformed signals. Unlike $CF$, $VCF$ weighting provides a ratio of coherent to the purely incoherent components of detected PA signals. The coherent components are considered as signal power contributions while incoherent components are mainly the noise power and artifacts generated in the sidelobes. Therefore, the presence of  $VCF$ suppresses the incoherent components without affecting the coherent part. In a nutshell, the proposed $VCF$ facilitates a fair assessment of signal-to-noise power under different noise levels. From Eq. \ref{eq:VCF_CF}, it is clear that (unlike $CF$) $VCF$ in combination with $DAS$ can suppress incoherent or noise contribution in the beamformed output.

Fluctuation in the delayed signals characterizes the performance of the reconstructed image quality. When PA signals are highly coherent in a particular imaging region, noise variation at that region gets lower. As a result, the measure of $\sigma$ is also lower and thus $VCF$ enhances the main lobe signal strength at that location. When incoherent components (noise and artifacts) are dominant, signal fluctuation or $\sigma$ increases. Therefore, the measurement of $VCF$ is reduced. In this way, $VCF$ acts as an adaptive feedback factor that suppresses the noise influence by lowering its value at the sidelobe region and preserving the coherent features at the mainlobe area.  It implies that $VCF$ is a statistical weight that can efficiently eliminate the noise components in the reconstructed signals. Therefore in $VCF$ beamformer, image quality metrics can be restored even when noise is high.

A generalized expression can be defined, whereby, a stabilization parameter $\beta$ under noise variation is introduced as \cite{wang2014snr}:
\begin{eqnarray}
(VCF)_\beta &=& \frac{<S>}{\sqrt{<S^{2}> - <S>^{2} + \beta <S>^{2}}}, \nonumber \\
        &=& \sqrt{\frac{CF}{1-CF(1-\beta)}}, \label{eq:modified_VCF} 
\end{eqnarray}
where, $S = [s_{1}[m, n], s_{2}[m, n], s_{3}[m, n],.....,s_{N}[m, n]]^{T}$ denotes $N\times 1$ matrix that represents an array of time delayed PA signal received by $N$ sensor elements of the ultrasound detector unit corresponding to PA source at a pre-specified point ($[m,n]$) in the reconstruction plane. $T$ refers to transpose. For the sake of simplicity, in Eq. \ref{eq:modified_VCF}, we drop all indices. The stability of the $VCF$ beamformer is characterized by the noise scaling factor ($\beta$). It gives a measure of coherent and incoherent components present in $VCF$ weighting. When noise scaling factor $\beta \sim 0$, $(VCF)_\beta$ (given in Eq. \ref{eq:modified_VCF}) is reduced to $VCF$ (Eq. \ref{eq:VCF_CF}). On the other hand, when $\beta$ is close to $1$, Eq. \ref{eq:modified_VCF} is reduced to Eq. \ref{eq:CF}, i.e., $(VCF)_\beta$ is reduced to $CF$.


\subsection{Impact of $VCF$ under source independent noise}
Assume that signals detected by individual channels of the (linear array) transducer unit are corrupted by a certain noise that is completely independent or uncorrelated to PA signals. This noise may be optical (due to fluctuation in incident (pulse) optical fluence), electrical (in receiving circuit), and acoustic (from ambient vibration) rather than the noise associated with the PA source. Assuming the noise ($\eta$) to be additive $-$ i.e., $S \rightarrow S + \eta$ $-$ we can deduce the noisy weighting factor (from Eq. \ref{eq:VCF_CF}) as:
\begin{eqnarray}
{\scriptstyle(VCF)_{Noise}} &=& {\scriptstyle\frac{<S+\eta>}{\sqrt{<(S + \eta)^{2}> - <(S+\eta)>^{2}}}}, \nonumber \\
    &=& {\scriptstyle\frac{<S> + <\eta>}{\sqrt{<S^{2}> + < \eta^{2}>+2<S><\eta> -<S>^2-<\eta>^2-2<S><\eta>}}}, \nonumber \\
    &=&{\scriptstyle\frac{<S>}{\sqrt{<S^{2}>-<S>^2+<\eta^{2}>}}}. \label{eq:VCF_noise}
\end{eqnarray}
Here, $\eta$ represents noise factor at the time of data acquisition in contrast to $\beta$ that corresponds to noise at the time of reconstruction. In Eq. \ref{eq:VCF_noise}, we take into account that noises are zero-mean, independent and uncorrelated so that all the average noise terms and signal-noise correlation terms present in Eq. \ref{eq:VCF_noise} are fixed to zero. From Eq. \ref{eq:CF} and Eq. \ref{eq:VCF_noise}, we can derive (detailed derivation is given in supplementary document (S)):
\begin{equation}
{\scriptstyle\frac{(VCF)_{Noise}}{VCF}} = {\scriptstyle\frac{VCF}{\sqrt{1+\frac{<N^{2}>}{<S^{2}>}\left(\frac{1}{1-CF}\right)}} = \frac{1}{\sqrt{1+\frac{P_N}{P_{S}}\left(\frac{1}{1-CF}\right)}}}.
\end{equation}
where, $P_S = <S^{2}>$ and $P_N = <N^{2}>$ represent signal power and noise power respectively. Again, $CF$ under external noise condition can also be defined in the similar way:
\begin{equation}
\frac{(CF)_{Noise}}{CF} = \frac{1}{\left(1+\frac{P_{N}}{P_{S}}\right)}.
\end{equation}
From Eq.(7) and (8), It can be noted that our proposed technique comprises additional information (in comparison to $CF$) under additive external noise. When the ratio of noise power to signal power increases, noise reduction capability of both weighting techniques ($CF$ and $VCF$) is degraded as usual. Still, at a particular noise level, $VCF$ encounters an additional incoherence part or target-related noise part ($1 - CF$) (see. Eq.7), that influences to eliminate sidelobes or grating lobe artifacts. As a result, the unwanted signals coming from sidelobes are reduced and it is manifested in our simulation and experimental results. In other words, our proposed technique outperforms other conventional ($CF$) methods.

Again, the capability of this beamforming technique can be described theoretically in terms of beamformer gain which is defined as the ratio of output powers for coherent component to the incoherent one \cite{hasegawa2020converting}. To derive the gain, we consider two different situations: (1) the detected PA signal consists of only coherent component (mainlobe signal) and in other case, (2) the received signals contain only incoherent components. Incoherent components are primarily random noise or out-of-focus signals (side lobes). In $VCF$ based technique, beamformer gain can be expressed as follows:
\begin{equation}
    A_{VCF} = \frac{N(N+(\beta - 1))}{\beta}\frac{p_{s}^{2}}{p_{n}^{2}},
    \label{eq:beamformer_gain}
\end{equation}
Beamformer gain in $CF$ based technique: 
\begin{equation}
    A_{CF} = N^{3}{\frac{p_{s}^{2}}{p_{n}^{2}}}.
    \label{cf_gain}
\end{equation}

where $N$ is the number of active transducer elements, $p_{s}^2$ is the output coherent power and $p_{n}^2$ is the noise power. The detailed derivation $CF$ and $VCF$ based beamformer gain is accomplished in supplementary document (S).  

In $CF$ based techniques (Eq. \ref{cf_gain}), beamformer gain is $\beta$ independent but for $VCF$, beamformer gain is $\beta$ dependent according to Eq. \ref{eq:beamformer_gain}. When $\beta$ is very small ($\sim 0$) $-$ beamformer gain is very high. Therefore, under low $\beta$ condition, $VCF$ performs significantly better than $CF$. But it is sensitive to the high value of  $\beta$. For high values of $\beta$, beamforming performance of $VCF$ is considerably reduced and can even go below $CF$.

\begin{figure}[htb!]
\centering
\includegraphics[scale=0.4]{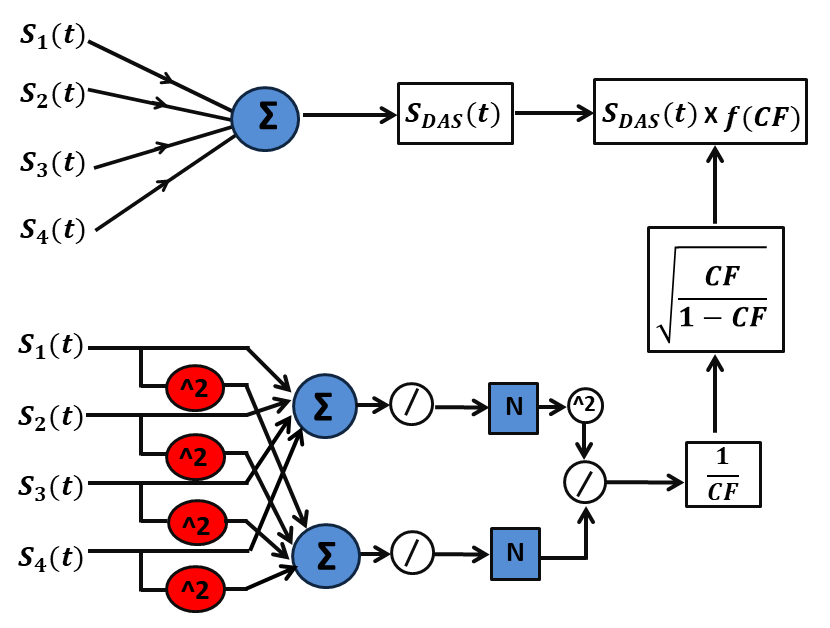}
\caption{The workflow diagram of the proposed variational coherence factor (VCF) based delay-and-sum ($DAS$) method, i.e., $DAS+VCF$.}
\label{work_flow.jpg}
\end{figure}

The workflow diagram of the proposed method is shown in Fig. \ref{work_flow.jpg}. In 1st step, the detected delayed signals ($4$ delayed signals are considered just for an example) are summed-up to compute $DAS$ beamformer then the resultant summed signal is squared and derived with the total number of elements ($N$). In 2nd step, all the delayed signals are individually squared, summed-up and divided by ($N$). Finally, step-$1$ and step-$2$ are followed to compute $CF$ and $VCF$. To speed up system execution and real-time performance, we have expressed $VCF$ as a function of the $CF$ (see Eq. \ref{eq:modified_VCF}) and thus, reduced the algorithm complexity. This algorithm enables to readily integrate our proposed method to the existing systems or algorithm workflows, and consequently, the improvised beamforming performance can be studied without any additional computational overhead. 

To obtain the beamformed images, $VCF$, as a weighting factor, is combined with $DAS$ beamformer for all the imaging points. For a particular point $[m,n]$ in the imaging domain, it can be expressed (similar to the conventional $CF$ based technique) as:
\begin{equation}
    S_{DAS + VCF}[m,n] = S_{DAS}[m,n]\times VCF[m,n].
\end{equation}
The additional weighting operation renders $DAS + VCF$ inherently slower than the native $DAS$ operation. However, the complexity is in par with the $CF$ based method as it is mentioned earlier in Sec II. 

\subsection{Quality assessment and validation}
In Sec. \ref{sec:simulation}, we provide a detailed account of numerical simulation and experimental results. We evaluated the image quality quantitatively and its improvements using various (image quality) metrics, viz., signal-to-noise-ratio ($SNR$), contrast ratio ($CR$) and generalized-contrast-to-noise-ratio ($gCNR$) \cite{kempski2020application} that can be mathematically, expressed as:
\begin{eqnarray}
    SNR &=& 20\log_{10}\left(\frac{\bar s_{in}}{\sigma_{n}}\right), \label{eq:SNR}\\
    CR &=& 20\log_{10}\left(\frac{\bar s_{in}}{{\bar s_{out}}}\right), \label{eq:CR} \\
    gCNR &=& 1 -\sum_{k=0}^{M-1}\text{min}[h_{i}(s_{k})h_{o}(s_{k})], \label{eq:gCNR}
\end{eqnarray}
where $\bar s_{in}$ is the mean of PA signal amplitude in target region of interest while $\bar s_{out}$ is the mean of PA signal strength in background (outside the target), $\sigma_{n}$ is the standard deviation of PA signal amplitude outside the target region, $h_{i}$ and $h_{o}$ are the histograms computed associated with target and background respectively \cite{kempski2020application}, $M$ is the total number of gray level values while $k$ represents gray level.) characterize probability of PA target detection \cite{kempski2020application}. 

Further, full-width-at-half-maximum ($FWHM$) of the acquired signal were estimated (from line plots) for all methods at different imaging depths to ascertain depth-resolved performance of the methods. Additionally, we carried out study on characterization of performance of the proposed reconstruction algorithm with different number of detector elements (64, 128 and 256). 

\section{Numerical simulation and Performance Evaluation}
\label{sec:simulation}

\subsection{Numerical simulation}
In this section, we present numerical simulation study to examine and compare the performance of our proposed algorithm with the existing algorithms (namely, $DAS$ and $DAS$ combined with $CF$). For the simulation study, we employed k-Wave MATLAB toolbox \cite{T_Cox, Souradip_P}, which is the most widely used numerical simulation tool in PA imaging study. In the simulation, we considered a homogeneous propagation medium, where, eight circular targets (as initial PA pressure sources) of same radius $0.1mm$ were situated. The targets were positioned along a vertical axis with a separation of $5mm$ apart from each other. The first target was $18mm$ away (along axial direction) from a linear transducer array sensor consisting of $N = 128$ elements that detect the generated PA signal from individual source. The sensor was kept along lateral direction. We assumed the speed of sound in the medium to be $1500m/s$ while the sampling frequency was $50MHz$. Envelope detection was performed using Hilbert transform in the beamformed signals to generate the final output images \cite{lutzweiler2015optoacoustic}. The entire algorithms and image reconstructions are implemented in MATLAB 2020a.

\begin{figure}[htb!]
\centering
\includegraphics[width=\columnwidth]{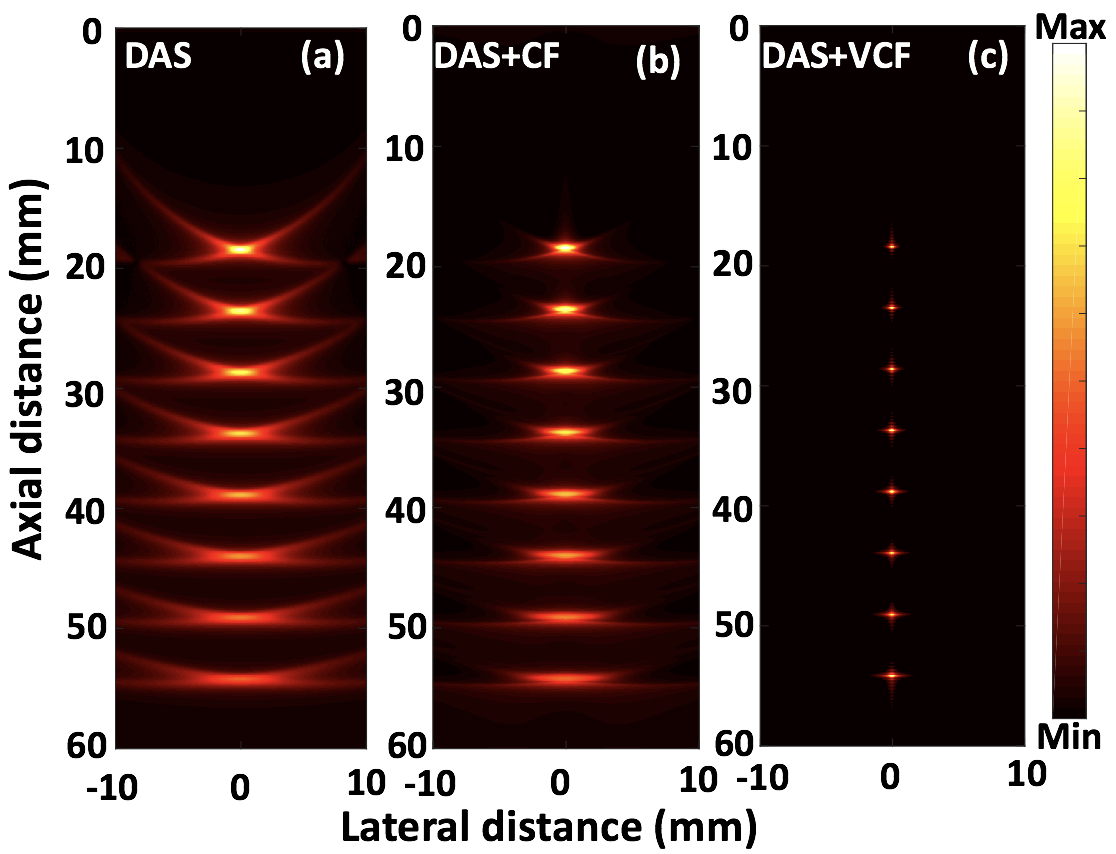}
\caption{Reconstructed images using the simulated detected data. (a) $DAS$, (b) $DAS + CF$, (c) $DAS + VCF$. A linear-array and point phantom were used for the numerical simulation study.}
\label{Point_tar.jpg}
\end{figure}

Reconstructed images of simulated phantom (containing of eight point targets) $-$ obtained by employing different reconstruction algorithms ($DAS$, $DAS+CF$ and $DAS+VCF$) $-$ are presented in Fig. \ref{Point_tar.jpg}. Figure \ref{Point_tar.jpg}(a) and \ref{Point_tar.jpg}(b) display the reconstructed images obtained by employing $DAS$ and $DAS$ combined with $CF$ (i.e., $(DAS + CF)$) respectively. Fig. \ref{Point_tar.jpg}(c) shows the reconstructed image obtained by using our proposed algorithm ($DAS + VCF$). In $DAS$ method, it is observed that beamforming image not only gives strong side lobes but also the provides poor spatial resolution. As the imaging depth is increased, the obtainable resolution is degraded both for $DAS$ and $DAS+CF$ algorithms. At depth beyond $33 mm$, targets are barely detectable as a point target (both in $DAS$ and $DAS+CF$). However, the side lobes are significantly reduced in the image (Fig.\ref{Point_tar.jpg}(b)) reconstructed by $CF$ combined with DAS beamforming while the sidelobes are disappeared in the reconstructed image (Fig. \ref{Point_tar.jpg}(c)) obtained by our proposed $DAS + VCF$. From the image (Fig. \ref{Point_tar.jpg}(c)), it is perceived that not only sidelobes and artifacts are completely suppressed but also targets at higher imaging depth are recovered and discernible. In the $(DAS + CF)$ beamformed image (Fig. \ref{Point_tar.jpg}(b)), recovery of targets at higher depth is improved (in comparison to $DAS$ method) but not satisfactory. With an increase of imaging depth, spatial resolution (both axial and lateral) obtainable with our proposed algorithm is also degraded $-$ but it is improved in comparison with that of the existing ($DAS$ and $DAS+CF$) algorithms $-$ which may be due to the loss of coherence of the beamformed PA signals arriving from higher imaging depths.

\subsection{Quantitative evaluation and image quality assessment}
\label{sec:quantitative_assess}
\begin{figure}[htb!]
\centering
\includegraphics[width = \columnwidth]{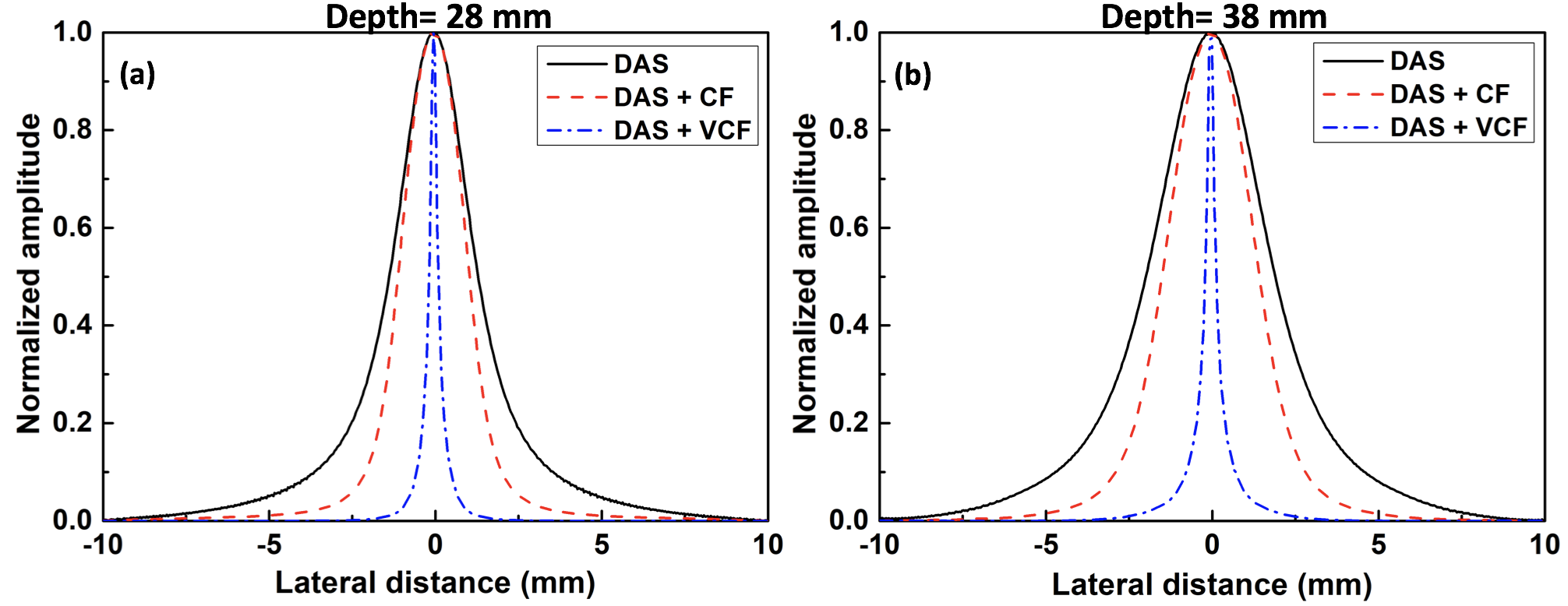}
\caption{Line plots that give variation of the reconstructed PA signal strengths along lateral direction across targets at the depth: (a) $28 mm$ and (b) $38 mm$.}
\label{lateral_pl_sim.png}
\end{figure}

To assess the performance of different reconstruction algorithms quantitatively, we adopt line plots that give variation of the reconstructed PA signal strength along a particular direction (profile across target of interest along lateral direction, in our present study). Fig. \ref{lateral_pl_sim.png}(a) and \ref{lateral_pl_sim.png}(b) depict variation of the reconstructed PA signal strength along lines (in lateral direction) passing through point targets at depth $28mm$ and $38mm$ respectively (for different reconstruction algorithms, namely, $DAS$, $DAS+CF$ and $DAS+VCF$). From the line plots, we estimated $FWHM$ to determine the lateral resolutions obtainable from the different methods and it is found to be: (1) $2.64 mm$ ($DAS$), $2.12 mm$ ($DAS+CF$) and $0.3 mm$ ($DAS+VCF$) (for target at $28mm$), and (2) $3.84 mm$ ($DAS$), $2.9 mm$ ($DAS+CF$) and $0.36 mm$ ($DAS+VCF$) (for target at $38mm$). The results demonstrate that our proposed $VCF + DAS$ method offers the narrowest possible main lobe width and consequently, the best possible spatial (lateral) resolution in comparison with conventional $DAS$ and $DAS + CF$ beamforming methods. A narrow mainlobe is associated with higher spatial resolution and it is determined by $FWHM$. 

\begin{figure}[htb!]
\centering
\includegraphics[width = \columnwidth]{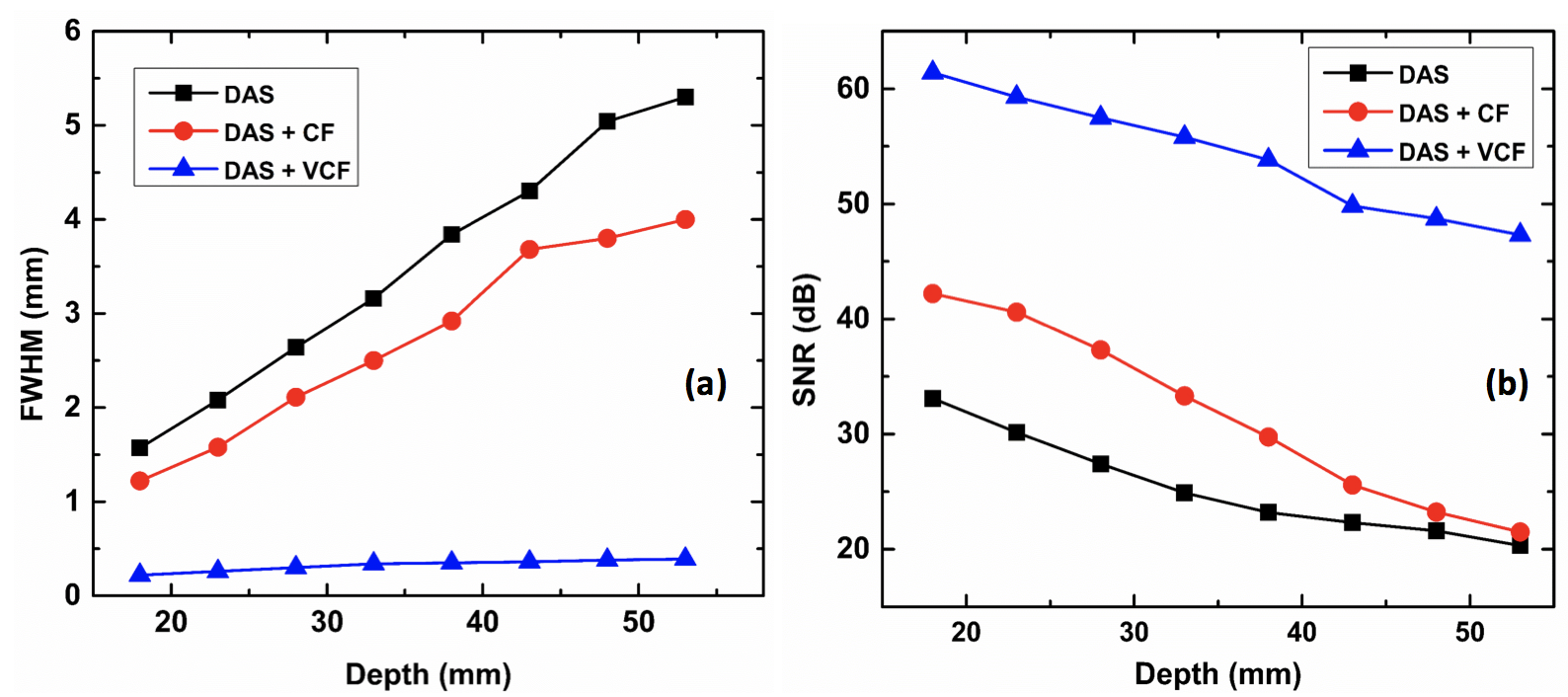}
\caption{Comparative performance of state-of-the art methods vs the proposed method ($DAS + VCF$): (a) $FWHM$ of the reconstructed point target at different imaging depths and (b) $SNR$ of the region of interest (targets) at different imaging depths.}
\label{FWHM_SNR.png}
\end{figure}

For quantitative analysis of obtainable spatial resolution with imaging depth, we estimated $FWHM$ for all of the points targets seated at different depths. Fig. \ref{FWHM_SNR.png}(a) gives the variation of the estimated $FWHM$ with imaging depth. From the figure, it is observed that $FWHM$ increases with imaging depth for $DAS$ and $DAS + CF$ while it is almost constant for $DAS + VCF$. This shows that image resolution $-$ achievable with the existing conventional methods ($DAS$ and $DAS+CF$) $-$ is significantly degraded with an increase in imaging depth (degrades by $3$\% at depth $\sim 5mm$). To be precise, for targets at depth $38 mm$, $FWHM$ for $DAS$, $DAS + CF$ and $DAS + VCF$ are found to be $3.84 mm$, $2.91 mm$ and $0.36mm$ respectively, i.e., our proposed algorithm significantly enhances the achievable spatial resolution in comparison with $DAS$ and $DAS + CF$ respectively. 

Using Eq. \ref{eq:SNR}, $SNR$ is calculated for all the target at different imaging depths. Figure. \ref{FWHM_SNR.png}(b) presents variation of $SNR$ with respect to the imaging depth. In this figure, one clearly observes that $SNR$ is significantly improved in comparison to that of $DAS$ and $DAS + CF$. In particular, for imaging target at depth of $38mm$, $SNR$ achievable with $DAS$, $DAS + CF$ and $DAS + VCF$ are $23.2 dB$, $29.74 dB$ and $48.8 dB$ respectively. This result convincingly demonstrates that our proposed algorithm outperforms the conventional algorithms providing a sharper beamformed images (Fig. \ref{FWHM_SNR.png}(a)). $SNR$ of the images reconstructed with $DAS + VCF$ method is approximately $20 dB$ higher in all imaging depths (see Fig. \ref{FWHM_SNR.png}(b)).

\begin{figure}[htb!]
\centering
\includegraphics[width = \columnwidth]{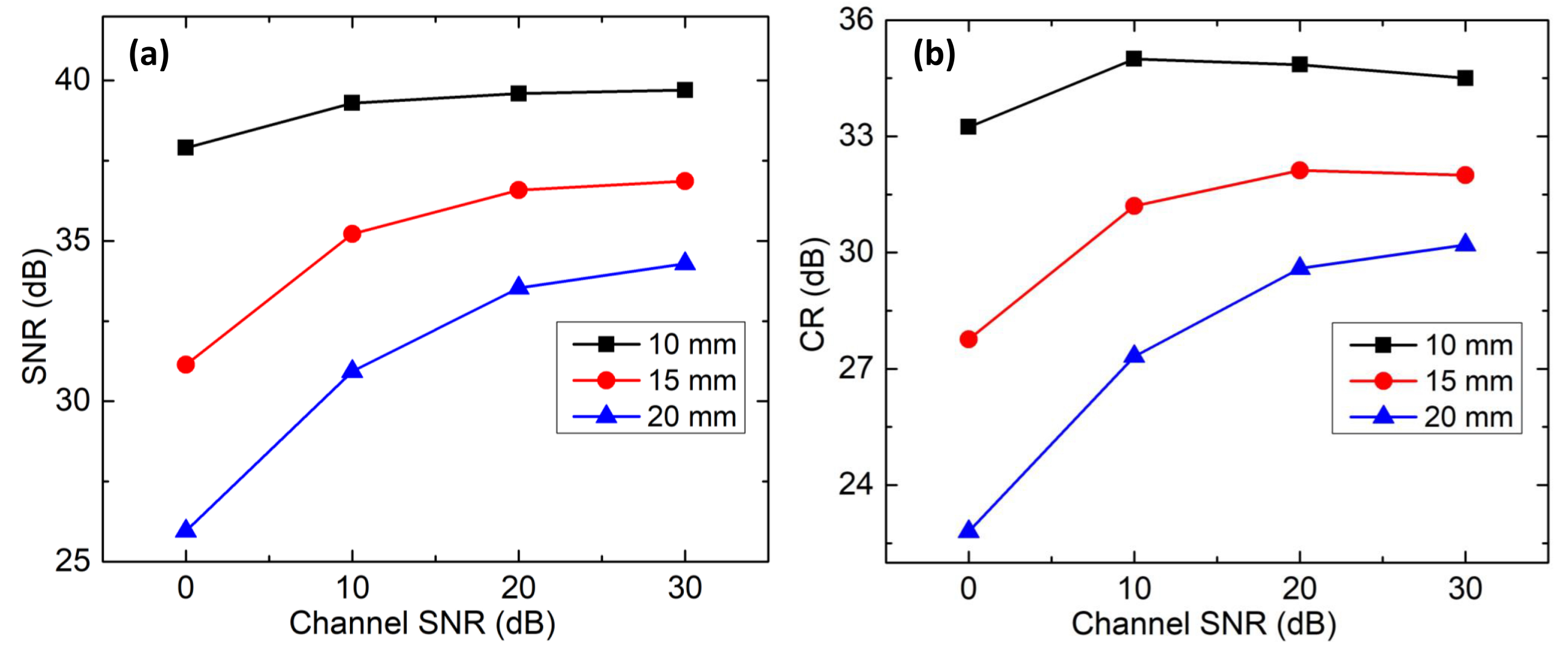}
\caption{$SNR$ (a) and $CR$ (b) variation as functions of channel $SNR$ for a simulated  target ($2 mm$ diameter) at $10 mm$, $15 mm$, and $20 mm$ imaging depths.}
\label{SNR_CR.png}
\end{figure}

We investigated the robustness of the proposed algorithm at various levels of (added) noise. We used a larger target (diameter $2 mm$) for the study. The channel $SNR$, that represents the noise levels on the desired channel, was added to the channel data using "addNoise.m" function in k-Wave toolbox. Fig. \ref{SNR_CR.png}(a) and Fig. \ref{SNR_CR.png}(b) show $SNR$ and $CR$ variances as channel $SNR$ is varied from $0 dB$ to $40 dB$ (for simulated targets being situated at different imaging depth ($10 mm$, $15 mm$ and $20 mm$). From Fig. \ref{SNR_CR.png}, it can be inferred that at higher channel $SNR$, $SNR$ and $CR$ of the reconstructed images decrease with an increase in imaging depth. 

\begin{figure}[htb!]
\centering
\includegraphics[width = \columnwidth]{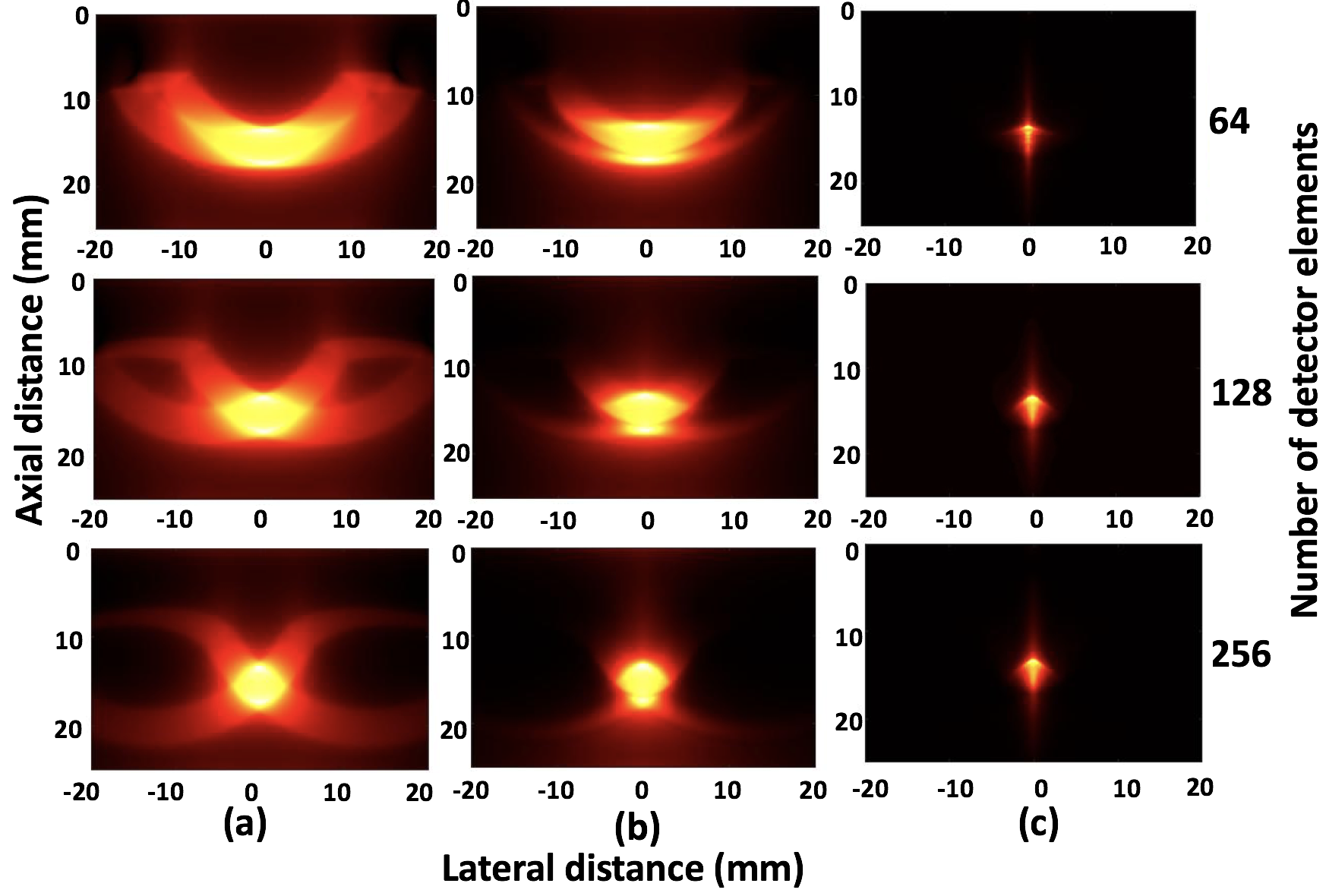}
\caption{Reconstruction of $2 mm$-diameter simulated target using $DAS$, $DAS + CF$ and $DAS + VCF$ techniques for different number of detector elements (64, 128 and 256).}
\label{Simulation- detector.png}
\end{figure}

\begin{figure}[htb!]
\centering
\includegraphics[scale=0.35]{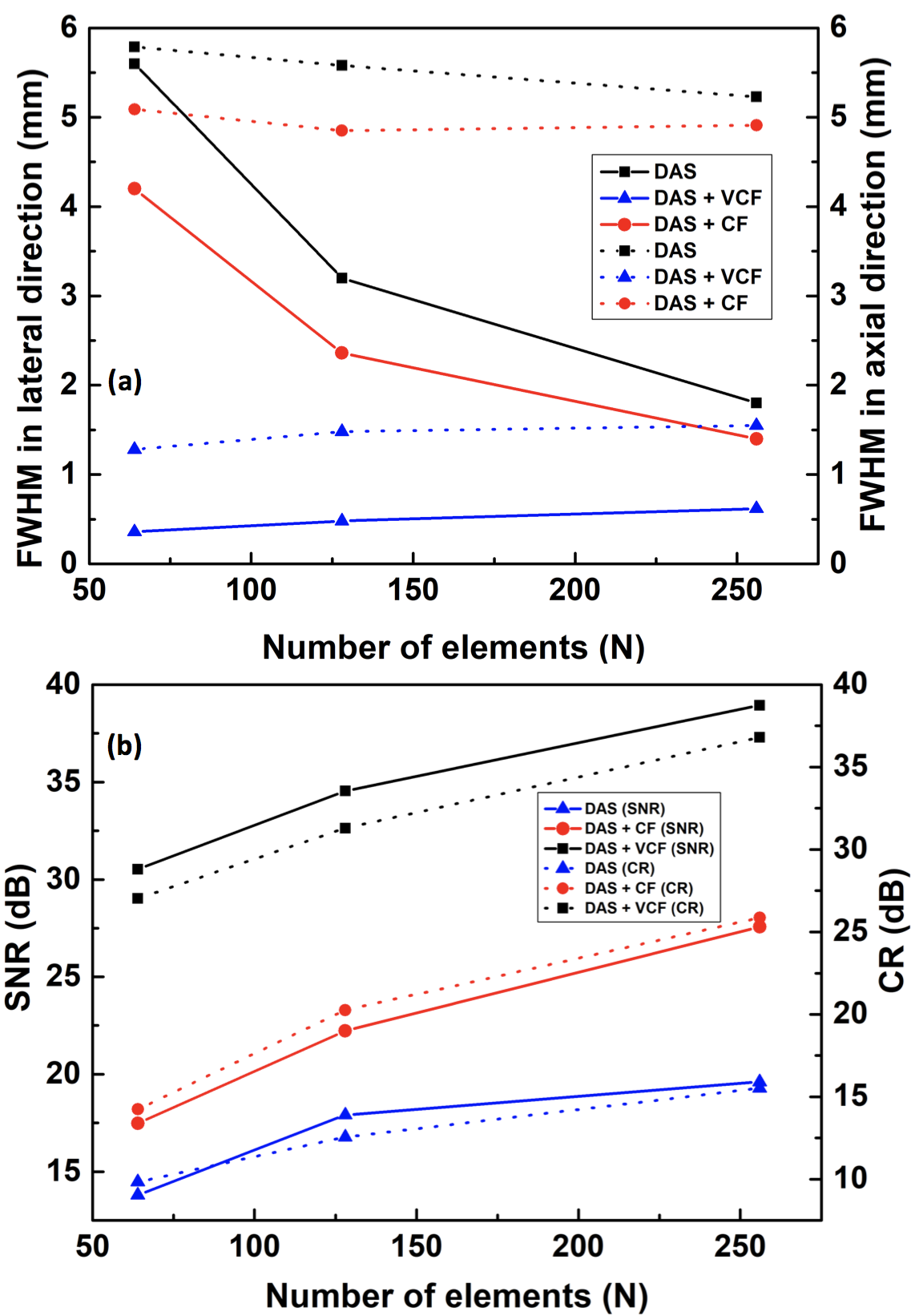}
\caption{Performance of beamformer under different sensor elements. (a) $FWHM$ along lateral (solid line) and axial direction (dotted line), (b) $SNR$ (solid line) and $CR$ (dotted line) are shown as functions of number of detector elements ($N = 64$, $128$ and $256$) for a $2 mm$ diameter simulated target at $15mm$ depth.}
\label{fig:Number_el_FWHM_SNR.png}
\end{figure}

Further, we undertook a study on improvement of reconstructed image quality as a function of the number of sensor elements on the ultrasound transducer unit. This study gives us a measure of the scalability of the proposed method \cite{Alexander_Dima, 7317595}. Fig. \ref{Simulation- detector.png} presents the beamformed images as a function of number of transducer elements (64, 128 and 256 elements) for different beamforming techniques. Quantitative analysis of performance of the beamforming algorithms $-$ employing $FWHM$, $SNR$ and $CR$ as statistical metric $-$ is shown in Fig. \ref{fig:Number_el_FWHM_SNR.png} (simulated target at $15mm$ imaging depth is chosen for the analysis). From Fig. \ref{Simulation- detector.png} and Fig. \ref{fig:Number_el_FWHM_SNR.png}, it is clearly noticeable that resolution performance of the state-of-the-art reconstruction algorithms ($DAS$ and $DAS + CF$) are highly dependent on the number of detector elements. However, in our proposed $DAS + VCF$ algorithm, image quality (image resolution and sidelobes) is less dependent on the number of detector and with 64 sensor elements, our proposed algorithm gives a satisfactory reconstructed image quality and can be further improved ($CR$, $SNR$) with additional detectors. This implies that our proposed algorithm facilitates to reduce the cost of PA imaging system dramatically $-$ which is known as one of the most expensive imaging devices that is primarily contributed from the acoustic sensor unit and its associated hardware (consisting of parallel amplifier and data acquisition system) $-$ without much compromise in obtainable image quality \cite{basak2020multiscale}. In addition, it reduces the data storage burden as well as greatly improves the speed of reconstruction \cite{meng2016high}. We investigated the beamforming performance under changing transducer frequencies (a short note of this observation is added in supplementary document (S)). The result exhibit that our proposed method provides better $SNR$ estimation than $DAS$ and $CF$ in all the frequency range. 

\section{Experimental results and validation}
\label{sec:expt_results}

\subsection{Experimental setup and imaging protocol}
\begin{figure}[htb!]
\centering
\includegraphics[width = \columnwidth]{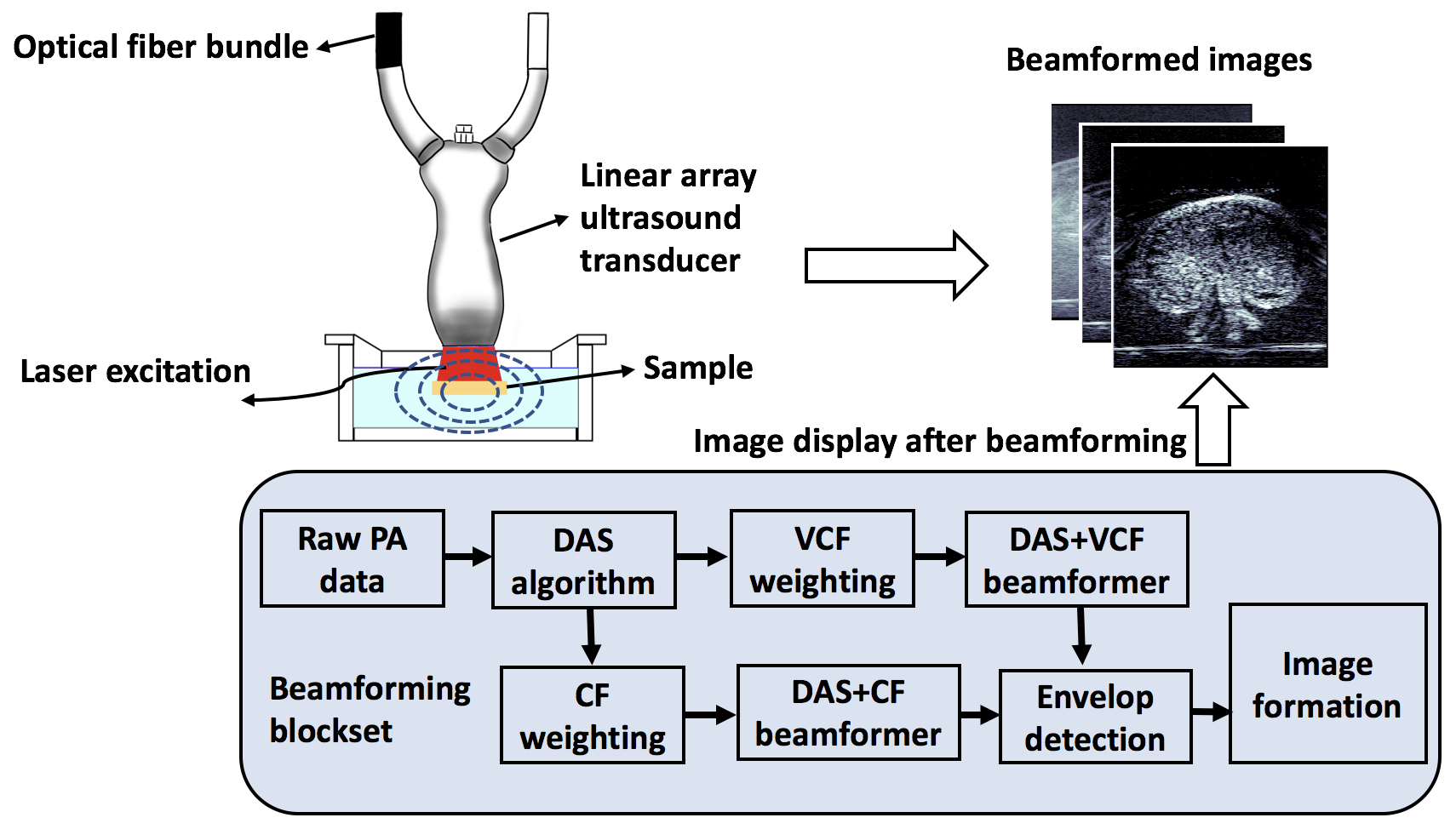}
\caption{Photoacoustic imaging system diagram and workflow for DAS-VCF beamforming. The beamforming blockset illustrate the methods employed for noise adaptive beamforming.}
\label{Picture_system.png}
\end{figure}

We used a hybrid PA-US system \cite{needles2013development} (Vevo LAZR-3100, Fujifilm VisualSonics) to acquire fused multimodal data in real-time. The laser sub-system generates $10ns$ pulsed laser beam at $680$-$970 nm$ for PA excitation with a $20Hz$ pulse repetition rate. A $256$ element high frequency custom designed linear array ultrasound probe with transmit frequency $15$-$30 MHz$ (center frequency $21 MHz$) and calibrated imaging resolution on $75 \mu m$ was used to detect the generated PA signals at a sampling rate of $84MHz$. For light delivery, a fibre bundle with split ends leading to two rectangular light bars mounted on each side of the transducer was used \cite{{needles2013development}}. Only one quarter of US array elements was activated per laser pulse, thus four pulse cycles were needed to reconstruct the entire cross-sectional image. Hence, the system had an effective frame rate of $5Hz$. Spatial resolutions were $165 \mu$m in the lateral direction and $75 \mu m$ in the axial direction was achieved for phantom and small animal experiments \cite{needles2013development}. The temperature was continuously monitored using a built-in temperature sensor and it was used as an initial estimate for speed of sound self-calibration \cite{mandal2014optimal}. A workflow diagram from PA signal acquisition to image reconstruction is presented in Fig. \ref{Picture_system.png}.

\subsection{Microsphere and Kidney phantom imaging}
We blocked five linearly spaced $120\mu M$ black polyethylene microspheres (Cospheric LLC, Santa Barbara, CA) in a circular Agar block of diameter $22.5mm$. The transducer was co-axially placed along the same optical path to provide suitable sheath illumination. The fluence variations were corrected by segmenting the surface of the imaging object and apply an finite volume based image correction algorithm \cite{7451214}.
\newline To demonstrate the performance of the methods on complex structures, we printed a  kidney (vascularized) image on $90 gsm$ paper using inkjet printer (shown in inset of Fig. \ref{phan_detector_kidney.png}(a)). The printed phantorm was blocked in tissue mimicking agar block and imaged at a pulse optical wavelength of $850 nm$.

\begin{figure}[htb!]
\centering
\includegraphics[width = \columnwidth]{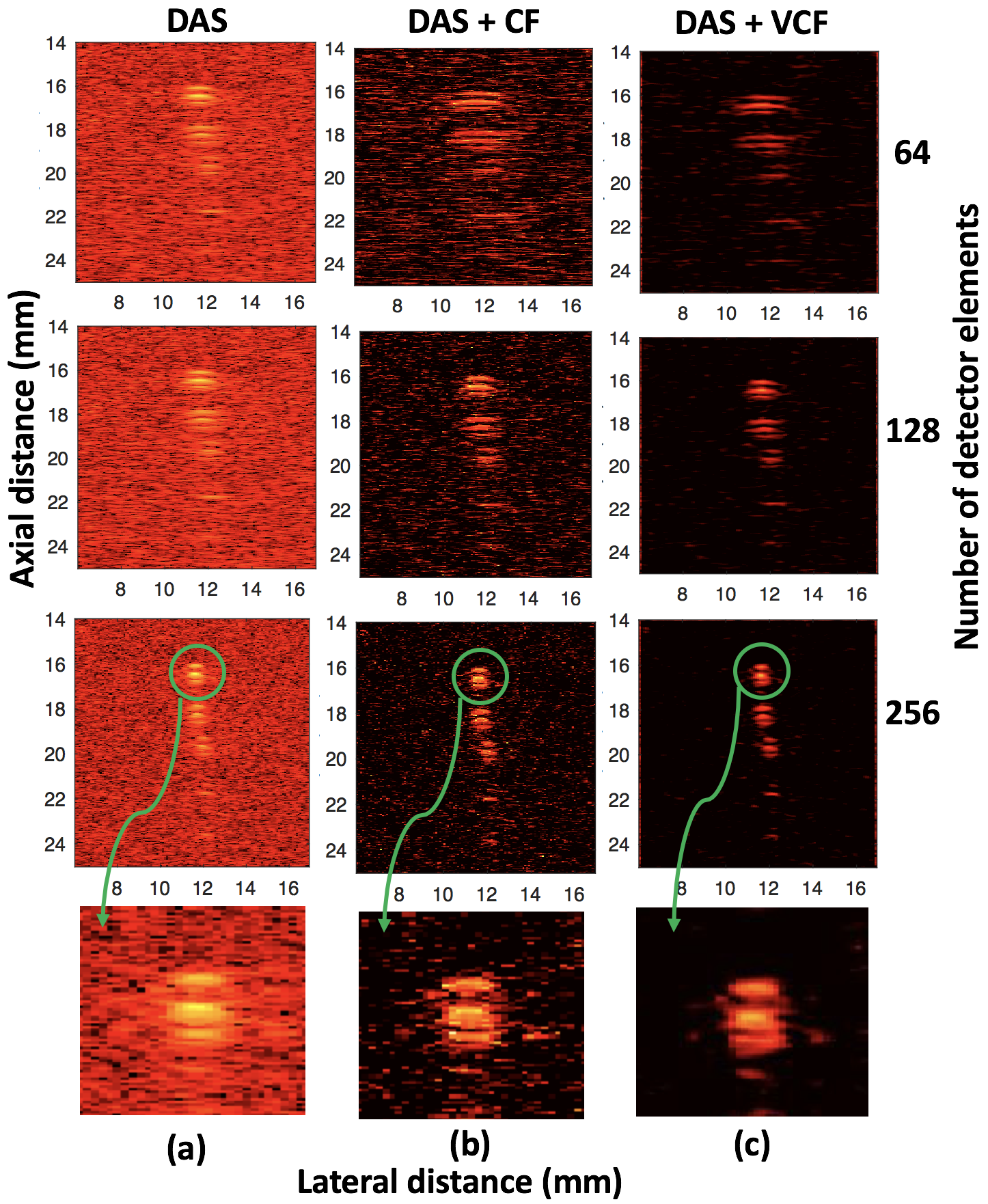}
\caption{Reconstruction of microsphere target using $DAS$, $DAS + CF$ and $DAS + VCF$ techniques for different number of detector elements ($64$, $128$ and $256$) is represented in 1st three row. Zoomed images of 1st microsphere (marked in green dotted circle) for 256 traducer elements (most bottom row) are shown using $DAS$, $DAS + CF$ and $DAS + VCF$ methods.}
\label{phan_detector.png}
\end{figure}

In Fig. \ref{phan_detector.png}, we show the reconstructions of the entire phantom with the change of transducer elements. The results demonstrate that $DAS$ is very noisy whereas $CF$ and $VCF$ based methods deliver significantly better performance. The improvements in $DAS + VCF$ is more apparent at the depths where the microspheres are differentiated from the noise unlike in other methods. The zoomed view of the first microsphere (closest to the transducer and light source) using $256$ elements further demonstrates the visual salience of the methods (most bottow row).

\begin{figure}[htb!]
\centering
\includegraphics[width = \columnwidth]{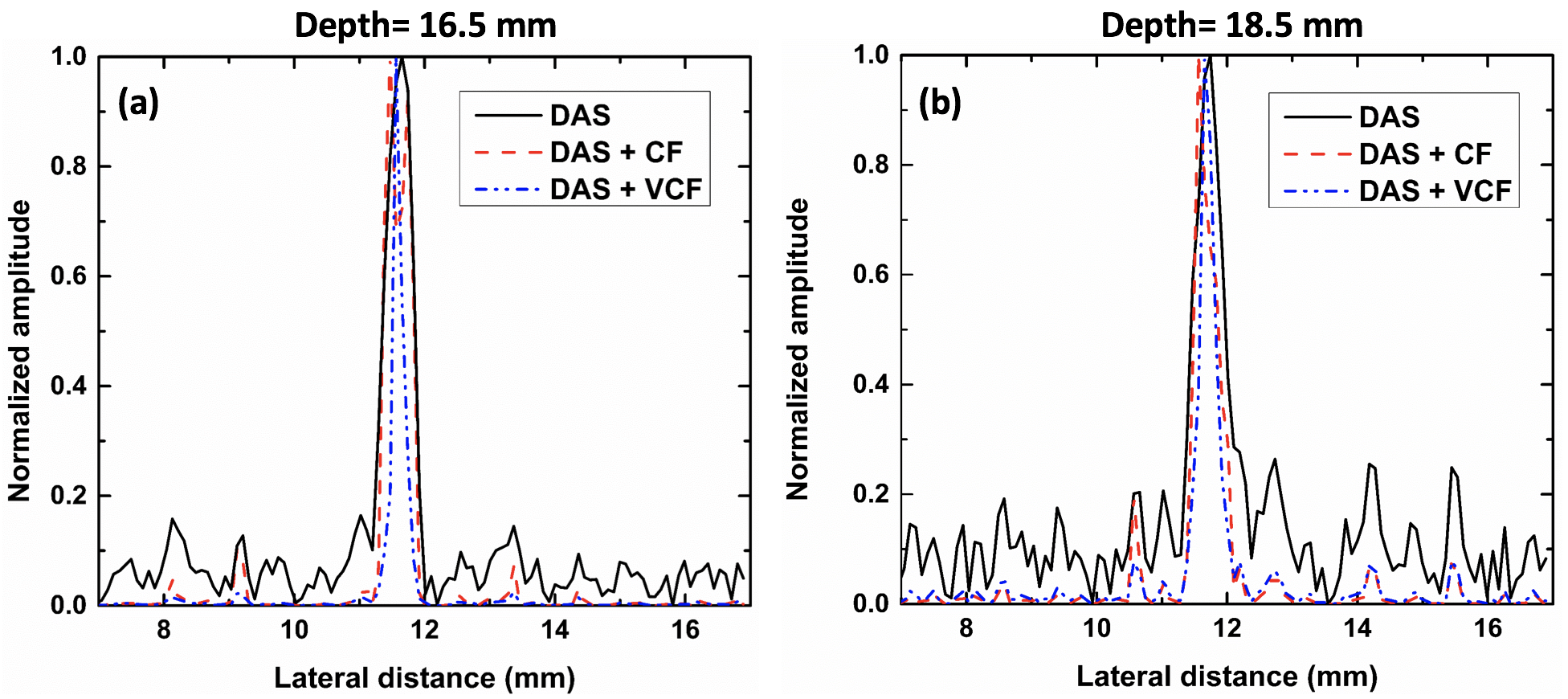}
\caption{Lateral variations of the 1st and 2nd reconstructed microspheres ($256$ elements transducer) in the Agar phantoms at two different depths (a) $16.5 mm$, (b) $18.5 mm$.}
\label{lateral_micro.png}
\end{figure}

The line plots of the the 1st and 2nd microspheres (at depths of $16.5mm$ and $18.5 mm$ from the surface) are shown in Fig. \ref{lateral_micro.png}. Similar to the results demonstrated in the simulations (Fig. \ref{lateral_pl_sim.png}), our proposed $DAS + VCF$ method performs better than the counterparts. 

\begin{table}[htb!]
\centering
\caption{Imaging performance of different beamforming methods at multiple depths}
\label{table 1}
\begin{tabular}{@{}|l|l|l|l|@{}}
\toprule
Depth ($mm$) & $DAS$ & $DAS+CF$ & $DAS+VCF$ \\ \midrule
\multicolumn{4}{|c|}{$FWHM$ ($mm$)} \\ \midrule
16.5 & 0.52 & 0.41 & 0.22 \\ \midrule
18.5 & 0.55 & 0.46 & 0.27 \\ \midrule
\multicolumn{4}{|c|}{$SNR$ ($dB$)} \\ \midrule
16.5 & 19.75 & 23.50 & 31.10 \\ \midrule
18.5 & 16.62 & 20.68 & 26.72 \\ \midrule
\multicolumn{4}{|c|}{$CR$ ($dB$)} \\ \midrule
16.5 & 14.80 & 26.78 & 29.70 \\ \midrule
18.5 & 11.23 & 22.64 & 23.32 \\ \midrule
\multicolumn{4}{|c|}{$gCNR$} \\ \midrule
16.5 & 0.92 & 0.94 & 0.98 \\ \midrule
18.5 & 0.88 & 0.89 & 0.93 \\ \bottomrule
\end{tabular}
\end{table}

\subsection{Quantitative performance evaluation}
To quantify the performance of the different beamforming algorithms, we ran a comparative test using the phantom model and the quantitative results are shown in Table \ref{table 1} for comparison. $FWHM$, $SNR$, $CR$ and $gCNR$ were computed for different imaging depths from the beamformed image data (see Table \ref{table 1}). Image quality was tested as a function of number of transducer elements (64, 128 and 256) using different beamforming techniques. The images were acquired in parallel acquisition mode, i.e., PA signals were captured on sequential quarter-segments of the 256 array elements for each laser pulse \cite{needles2013development}. Thus, we sub-sampled the data in each quarter-segment and reconstructed the subset of the data to obtain the results for 64 and 128 elements. For 256 elements, the complete dataset was used to reconstruct the image. Fig. \ref{phan_detector.png} represents the improvement of image quality for the increasing number of transducer elements. The results conclusively demonstrate that the proposed method outperforms the existing algorithms in terms of all the selected parameters and it is quite consistent to the results obtained from the simulation study presented in Sec. \ref{sec:simulation}. \\

\begin{figure}[htb!]
\centering
\includegraphics[width = \columnwidth]{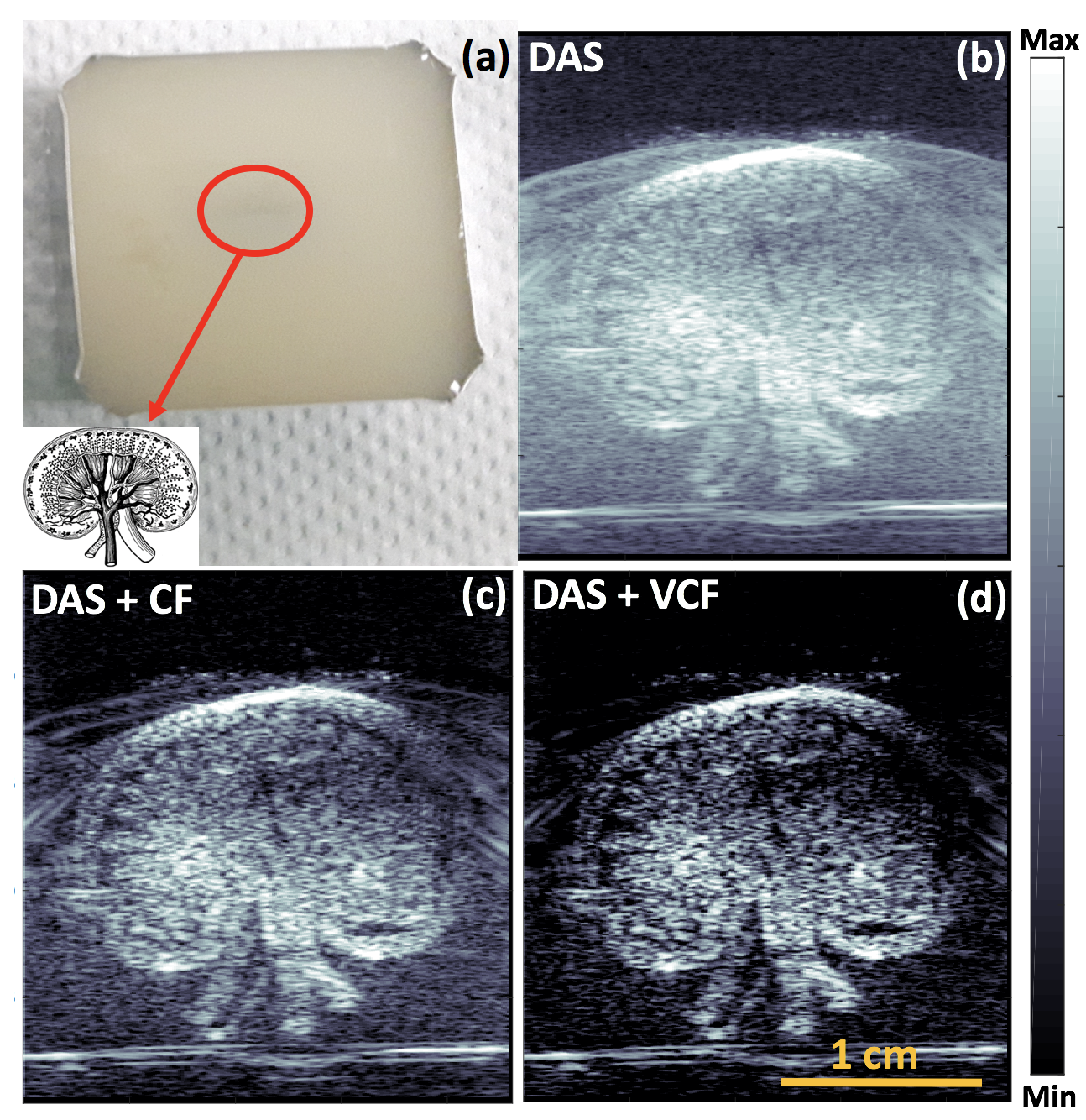}
\caption{Reconstruction of  a complex phantom (printed kidney with internal vascular structures illustrated) using $DAS$, $DAS + CF$ and $DAS + VCF$ techniques.The proposed algorithm shows reduced background noise, good structural recovery and high signal fidelity in deeper structures (scale: 1 cm).}
\label{phan_detector_kidney.png}
\end{figure}

Reconstructed image using $DAS$, $DAS+CF$ and $DAS$ combined with $VCF$ for kidney phantom is exhibited in Fig. \ref{phan_detector_kidney.png}. As it can be easily observed, The reconstructed image using $DAS$ beamforming (Fig. \ref{phan_detector_kidney.png}(b)) suffers from high levels of artifacts due to the non-adaptiveness and blindness of this algorithm, especially in the boundary region of the kidney phantom. Image quality gradually improved when it is combined with adaptive weighting factors ($CF$ and $VCF$) in Fig. \ref{phan_detector_kidney.png}(c) and \ref{phan_detector_kidney.png}(d). The overall boundary of the kidney phantom is nicely visible from its background and sidelobes, artifacts are also highly reduced in our proposed technique (Fig. \ref{phan_detector_kidney.png}(d)). All the quantitative assessment of the image quality metrics (Table \ref{table 2}) suggest that $VCF$ based beamformer has a great potentiality in PA image improvement.

\begin{table}[htb!]
\centering
\caption{Imaging performance of different beamforming methods for kidney phantom.}
\label{table 2}
\begin{tabular}{@{}|l|l|l|l|@{}}
\toprule
Depth ($mm$) & $DAS$ & $DAS+CF$ & $DAS+VCF$ \\ \midrule
\multicolumn{4}{|c|}{$SNR$} \\ \midrule
17.5 & 14.20 & 18.93 & 23.10 \\ \midrule
\multicolumn{4}{|c|}{$CR$} \\ \midrule
17.5 & 8.67 & 28.68 & 29.20 \\ \midrule
\multicolumn{4}{|c|}{$gCNR$} \\ \midrule
17.5 & 0.69 & 0.71 & 0.83 \\ \bottomrule
\end{tabular}
\end{table}
Real time performance is critical to clinical translation of PA imaging methods, thus we aimed at optimizing the imaging speed. The computational complexity of VCF is mainly determined by mean and square mean operation of the detected data, which is $O(N)$. The beamforming algorithms were bench-marked on a portable laptop computer with Intel Core i7-8565 CPU at 1.80GHz and 16GB RAM to compare the computing performance. The code was partially parallelized and used a parallel computing toolbox to efficiently compute the noise variance and delays for beamforming. The reconstruction time was summed over a stack of $33$ frames is shown in Table \ref{table 3}. The processing time using $VCF$ is marginally slower than $DAS$ beamforming, but, the time-frame is in same or faster than $DAS + CF$ beamforming. No GPU acceleration was performed in the current experiments.

\begin{table}[htb!]
\caption{Reconstruction time of different beamforming methods for microsphere phantom}
\label{table 3}
\begin{center}
\begin{tabular}{|c||c||c||c|}
\hline
\multicolumn{4}{|c|}{Reconstruction time ($ms$)} \\
\hline
Mode & $DAS$ & $DAS+CF$ & $DAS+VCF$\\
\hline
Experimental & 868.87 & 879.21 & 876.44 \\
\hline
Complexity & $N$ & $3N+2$ & $3N+5$ \\
\hline
\end{tabular}
\end{center}
\end{table}
The results indicate that the $VCF$ technique computationally relatively inexpensive and it can be implemented in a real-time PA imaging application. Given the recent advance of hardware platform, we are confident that our method can be scale to real time operation in near future.

\section{Discussion}
\label{sec:discussion}
We introduced a novel weighting factor based on mean-to-standard-deviation-ratio to enhance the performance of PA imaging. The major benefits of the proposed $VCF$ method are achieving higher image resolution, sidelobe suppression and contrast improvement. Simulation and experimental results argue that the proposed method improves image resolution, contrast, and $SNR$ compared with the traditional $DAS$ image reconstruction. It reduces the noise and provides high quality images. $DAS$ reconstructions are plagued with low quality image having poor resolution and strong sidelobes due to its blindness and non-adaptiveness. $CF$ is an adaptive weighting method that can be used with $DAS$ or $MV$ beamformers, for sidelobes reduction. However, the proposed method outperforms existing $CF$ beamforming method in terms of improving image quality and speed of execution. $DAS + CF$ is a weighting method which can be used in conjunction with any beamforming algorithm ($DAS$ was used in this paper) thus extending the applicability of the method to a wide range of scenarios. The reconstructed images (Fig. \ref{Point_tar.jpg}, Fig. \ref{Simulation- detector.png} and Fig. \ref{phan_detector_kidney.png}) show that $VCF$ outperforms $CF$ method. As shown in Fig. \ref{Point_tar.jpg}, the point targets are much more distinguishable and detectable at higher imaging depth using our technique. It can be seen in the lateral variations in Fig. \ref{lateral_pl_sim.png} and \ref{lateral_micro.png}, the width of mainlobe has decreased due to the use of $VCF$. The proposed method shows the superiority when the targets are present at high depths of imaging, shown in Fig. \ref{FWHM_SNR.png}. As shown in Fig. \ref{Point_tar.jpg}, for the targets located at the depths of $43$ to $53 mm$, sidelobes and artifacts are nicely reduced. Quantitative evaluation of image quality metrics under various noise conditions are presented in Fig. 5 and 7, which shows the good stability of the algorithm under different noise floors and the number of transducer elements. Performance of $SNR$ and $CR$ with the variation of channel noises and number of transducers is evident that $VCF$ beamformer provides better image quality metrics not only w.r.t. to the conventional techniques but also under varying noise conditions. Both simulation (Fig. \ref{Simulation- detector.png}) and phantom studies (Fig. \ref{phan_detector.png}) demonstrates that the proposed algorithm gives satisfactory image quality for simulated phantom with lesser number of sensor element (64/128 elements) in comparison to that of the standard algorithms using higher number of sensor elements (256 elements). The qualitative and quantitative results obtained using tissue mimicking phantoms corroborated the fact that $DAS + VCF$ leads to contrast improvement, noise reduction, and narrower mainlobe as compared to similar state-of-the-art methods.

The unique capability of the algorithm to improve image quality with low hardware or computational overhead can remarkably reduce the cost of PA imaging system by reducing the requirements of detection channels/elements and associated hardware including DAQ system. The reduction of acquisition channels can further optimize the computation performance of the PA systems. As it is outlined in Table III, the run-time of the algorithm for full data-stack (all elements) in comparable with other methods can be used for video rate imaging (on suitable hardware platforms). The better performance of this algorithm comes at the slight increment of computational time, which could be a minor concern for real-time implementations. Reducing the computation burden will enable accelerating performance that might be useful for perfusion and functional imaging studies \cite{ermolayev2016simultaneous}. With the recent developments of hardware systems, GPU and parallel computing, the proposed algorithm can be significantly accelerated and leading towards adoption in clinical imaging \cite{steinberg2019photoacoustic}. Recent clinical trial in PA imaging has shown its efficacy in oncological imaging, lymph node mapping and chronic inflammatory diseases using hand-held linear array probe \cite{stoffels2019assessment, knieling2017multispectral, kothapalli2019simultaneous}. Since the proposed method also focuses on performance improvement in similar photoacoustic probe, we believe that our contribution will be useful further towards the clinical translation of the modality. 
In recent time, Light emitting diode (LED) based PA imaging system are being increasingly used due to its low cost and versatility \cite{zhu2018light, upputuri2018fast, basak2020multiscale}. The proposed method can be easily adapted to such LED based systems. Owing to the noise adaptive nature of the algorithm it will be useful to address the limitation of the LED based systems which are inherently noisy and has significantly lower $SNR$ than traditional optical resonator based laser systems. This method can also be a useful technique to improve the strain-based noisy reconstructed images in ultrasound and photoacoustic elastography \cite{gao2019learning, mandal2019determining}. 

\section{Conclusion}
\label{sec:conclusion}
In this paper, we developed a new weighting procedure for PA imaging to enhance image quality and overcome the limitation of $DAS$ beamformer. The proposed method was implemented and validated using simulated and experimental data. All the simulation results showed that $VCF$ has excellent performance in resolution, $SNR$ and sidelobe reduction. The experimental results confirmed the effectiveness of this proposed method at improving resolution, $SNR$ and $CR$ quantitatively. The computational burden of $DAS + VCF$ method is low due to its simple implementation. Therefore, we believe that our method is well suited for cost effective real time PA imaging applications and it can be integrated into clinical photoacoustic systems for attaining better image quality and enhanced clinical outcomes.

\section*{Acknowledgment}
The authors would like to thank Dr. Dorde Komljenovic and Prof. Mark E. Ladd (DKFZ Heidelberg) for their support towards the work. 

\bibliographystyle{ieeetr}
\bibliography{bibliography.bib}

\end{document}


\maketitle

\section{Supplementary (S)}

\subsection*{Derivation of $VCF$ weighting}
We define variational coherence factor ($VF$) as the ratio of mean ($\mu$) and standard deviation ($\sigma$) of the detected delayed signals. Mathematically, we can write as:
\begin{eqnarray}
 VCF[m,n]&=&  \frac{\mu[m,n]}{\sigma[m,n]}, \nonumber \\
                         &=&  \frac{\frac{1}{N}\sum_{i=1}^{N}s_i [k + \Delta_i [m,n] ] }{\sqrt{\frac{1}{N} \sum_{i=1}^{N} \left (s_i [k + \Delta_i [m,n] ] - \frac{1}{N}\sum_{i=1}^{N}s_i [k + \Delta_i [m,n] ] \right ) ^{2}}}, \nonumber \\       &=&  \frac{<S [k + \Delta [m,n] ]>}{\sqrt{\frac{1}{N} \sum_{i=1}^{N} \left (s_i [k + \Delta_i [m,n] ] - <S [k + \Delta [m,n] ]> \right ) ^{2}}}, \nonumber \\
                         &=&  \frac{<s [k + \Delta [m,n] ]>}{\sqrt{\frac{1}{N} \sum_{i=1}^{N} \left (s_i^{2} [k + \Delta_i [m,n] ] + <S [k + \Delta [m,n] ]>^{2} - 2 s_i [k + \Delta_i [m,n] ] <S [k + \Delta [m,n] ]> \right )}}, \nonumber \\
                         &=&  \frac{<S [k + \Delta [m,n] ]>}{\sqrt{<S^{2} [k + \Delta [m,n] ]> - <S [k + \Delta [m,n] ]>^{2}}}, \nonumber \\
                         &=&  \frac{1}{\sqrt{\frac{<S^{2} [k + \Delta [m,n] ]>}{<S [k + \Delta [m,n] ]>^{2}}-1}}, \nonumber \\
                         &=&  \frac{1}{\sqrt{\frac{1}{CF}-1}}, \nonumber \\
                         &=&  \sqrt{\frac{CF}{1-CF}}, \nonumber \\
                         &\equiv&  f(CF).
\end{eqnarray}

\subsection*{Derivation of $VCF$ weighting under source independent noise}

\begin{eqnarray}
(VCF)_{Noise} &=& \frac{<S+\eta>}{\sqrt{<(S + \eta)^{2}> - <(S+\eta)>^{2}}}, \nonumber \\
    &=& \frac{<S> + <\eta>}{\sqrt{<S^{2}> + < \eta^{2}>+2<S><\eta> -<S>^2-<\eta>^2-2<S><\eta>}}, \nonumber \\
    &=&\frac{<S>}{\sqrt{<S^{2}>-<S>^2+<\eta^{2}>}} \nonumber \\
    &=& \frac{<S>}{\sqrt{<S^{2}>-<S>^2}}\frac{1}{\sqrt{1+\frac{<\eta^{2}>}{<S^{2}>-<S>^2}}}\nonumber \\ 
    &=& \frac{VCF}{\sqrt{1+\frac{<\eta^{2}>}{<S^{2}>}\frac{1}{1-CF}}}.
\end{eqnarray}

\subsection*{$VCF$ based beamformer gain}
The expected power ($P_{VCF}$) of $VCF$ based beamforming technique while combine with $DAS$ under noise scaling parameter (Eq. 5 in the main text) is expressed as:
\begin{eqnarray}
P_{VCF} &=& \left |S_{DAS+VCF}\right |^{2}, \nonumber \\
        &=& E\left[((VCF)_{\beta})^2\right].E\left[(S_{DAS})^2\right],\\
        &=& \frac{E\left [\left |\frac{1}{N}\sum_{i}^{N}s_{i} \right |^{2} \right ].E \left [\left |\frac{1}{N}\sum_{i}^{N}s_{i} \right |^{2} \right]}
        {E \left [\frac{1}{N}\sum_{i}^{N}s_{i}^{2} \right] - (1-\beta)E \left [\left |\frac{1}{N}\sum_{i}^{N}s_{i} \right |^{2} \right]},
\end{eqnarray}
where $E[.]$ denotes the expectation operation.  

Consider a situation, when PA signals contain only a coherent component, that is, signals are mostly from the mainlobe or target region or focus region. The power of the $VCF$ based technique is expressed as follows:
\begin{equation}
    P_{VCF} = \frac{p_{s}^{4}}{p_{s}^{2}-(1-\beta)p_{s}^{2}} = \frac{p_{s}^2}{\beta},
    \label{vcf_sig}
\end{equation}
where, $E\left [ \left|\frac{1}{N}\sum_{i}^{N}s_{i} \right|^{2}\right ] = E\left [\left |\frac{1}{N}\sum_{i}^{N}s_{i}^{2}\right | \right] = p_{s}^2$ (PA signal contains only a coherent component).

In the other extreme situation where the signals are composed of noise ($\eta_{i}$) only. The output beamformed signal will be:
\begin{eqnarray}
 E\left [\left |\frac{1}{N}\sum_{i}^{N}s_{i}^{2}\right | \right] &=&  E \left [ \left|\frac{1}{N}\sum_{i}^{N}n_{i}^{2} \right | \right] = p_{n}^{2}, \nonumber \\ 
  E \left[ \left|\frac{1}{N}\sum_{i}^{N}s_{i}\right|^{2}\right] &=& E\left[\left|\frac{1}{N}\sum_{i}^{N}n_{i}\right|^{2}\right] = \frac{p_{n}^{2}}{N}.
\end{eqnarray}

The expected value of the zero-mean random noise $n_i$ will be zero if the number of detection is infinite. As we are having finite number of detection $N$, noise power will be suppressed by $\frac{1}{N}$. 

If the received PA signal $s_{i}$ contains only noise term $(n_{i})$ (incoherent signal), $P_{VCF}$ is expressed as:
\begin{eqnarray}
P_{VCF} &=& \frac{E\left [\left |\frac{1}{N}\sum_{i}^{N}n_{i}\right|^{2}\right].E \left [\left |\frac{1}{N}\sum_{i}^{N}n_{i}\right |^{2}\right ]}{E\left [\frac{1}{N}\sum_{i}^{N}n_{i}^{2}\right] - (1-\beta)E\left [\left |\frac{1}{N}\sum_{i}^{N}n_{i}\right|^{2}\right]}, \nonumber \\
        &=& \frac{\frac{p_{n}^{2}}{N}\frac{p_{n}^{2}}{N}}{p_{n}^{2} - \frac{p_{n}^{2}}{N} +\beta.\frac{p_{n}^{2}}{N}}, \nonumber \\ 
        &=& \frac{p_{n}^{2}}{N(N+(\beta - 1)))}.
        \label{vcf_noise}
\end{eqnarray}

Therefore, beamformer gain $(A_{VCF})$ can be obtained by dividing Eq. \ref{vcf_sig} and \ref{vcf_noise},
\begin{equation}
    A_{VCF} = \frac{N(N+(\beta - 1))}{\beta}\frac{p_{s}^{2}}{p_{n}^{2}}.
\end{equation}

In the similar way, $CF$ based beamformer gain can be derived as:
\begin{equation}
    A_{CF} = N^{3}{\frac{p_{s}^{2}}{p_{n}^{2}}}.
\end{equation}

\subsection*{SNR at different channel noise and differnet central frequency of the transducer}
\begin{figure}[htb!]
\centering
 \includegraphics[scale=0.5]{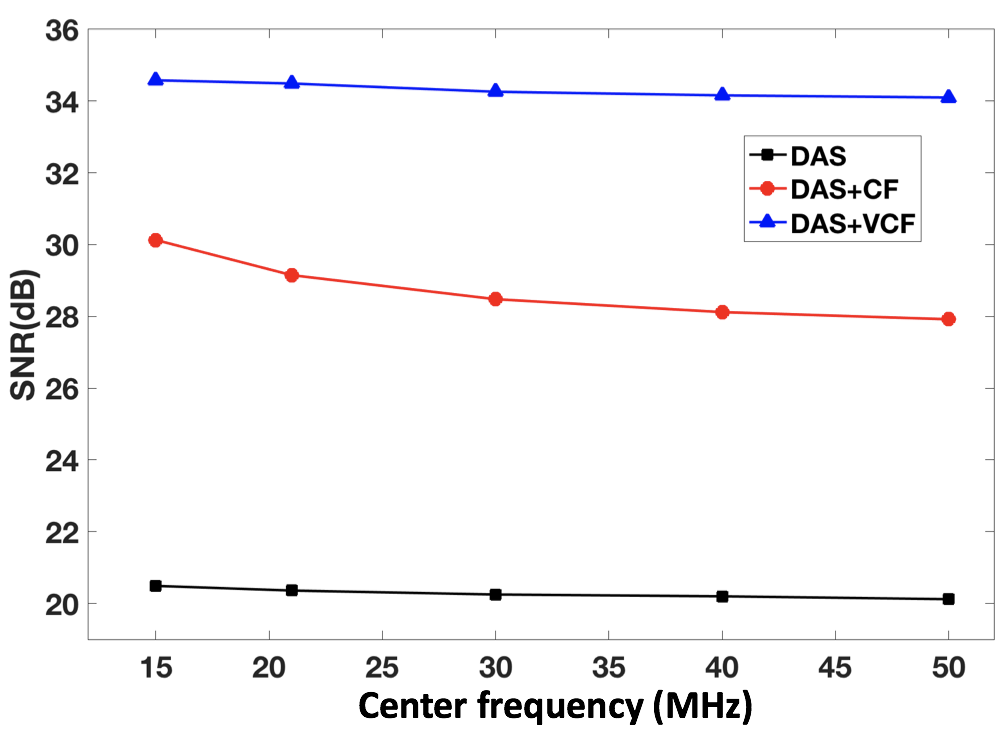}

{{\bf{Fig.S1:}}  Variation of $SNR$ with different central frequency of the transducer for a simulated target ($2mm$ diameter) at $15mm$ imaging depth.}
\label{SNR_noise}
\end{figure} 

To demonstrate the performance of the proposed beamforming algorithm under different frequency responses, we evaluate SNR at various central frequencies of the transducer. The frequency was varied from $15$ to $50Mhz$ using numerical simulations. From Fig. S1, It is observed that change of SNR with transducer center frequency is not significant with our proposed method. Compared with DAS and CF methods, our proposed method outperforms in all the frequency range and provides better SNR estimation.